\documentstyle[12pt]{article}
\input epsf
\begin{document}
\newcommand{\lsi}{\,\raisebox{-0.13cm}{$\stackrel{\textstyle<}
{\textstyle\sim}$}\,}
\newcommand{\gsi}{\,\raisebox{-0.13cm}{$\stackrel{\textstyle>}
{\textstyle\sim}$}\,}

\rightline{RU-96-06}
\rightline{hep-ph/969602334}

\baselineskip=18pt
\vskip 0.7in
\begin{center}
{\bf \LARGE Signatures for Squarks\\
in the Light Gaugino Scenario}\\ 
\vspace*{0.4in}
{\large Glennys R. Farrar}\footnote{Research supported in part by
NSF-PHY-94-23002} \\
\vspace{.1in}
{\it Department of Physics and Astronomy \\ Rutgers University,
Piscataway, NJ 08855, USA}\\
\end{center}
\vspace*{0.2in}
\vskip  0.9in  

{\bf Abstract:} When the gluino is light and long lived, missing energy
is a poor signature for both squarks and gluinos.  Instead, squark pair
production leads to events with $\ge 4$ jets.  If a chargino can decay
to squark and quark, missing energy is also a poor signature for the
chargino.  Properties of 4-jet events originating from squarks are
discussed.  ALEPH's preliminary report of an excess of 4-jet
events, with a peak in total dijet mass of 109 GeV, is analyzed in
terms of $S_q S_q^*$ and chargino pair production.

\thispagestyle{empty} 
\newpage 
\addtocounter{page}{-1} 
\newpage

Motivations are accumulating for believing the gluino and photino may
be light.  This is inevitable as long as SUSY-breaking does not arise
from gauge singlet vevs and is transmitted to ordinary particles by
exchange of very heavy states. In this case gaugino masses and other
dimension-3 SUSY-breaking operators are suppressed by a factor $M^{-1}$
compared to squark masses and other dimension-2 SUSY-breaking operators.
This has a number of attractive concomittants:  
\begin{itemize}
\item  There is no SUSY CP problem\cite{f:99}.
\item Gluino and photino masses are calculable in terms of $\mu$, $tan
\beta$, and squark and Higgs masses.  Given constraints on these parameters,
gluino and photino masses are $\le O(1)$ GeV.  The mass of the lightest
gluino-containing hadron (the gluino-gluon bound state called $R^0$) is $\sim
1.3-2.2$ GeV\cite{f:99}.  A chargino is lighter than the
$W$\footnote{Unless $\mu {\,\raisebox{-0.13cm}{$\stackrel{\textstyle>}
{\textstyle\sim}$}\,} $ few TeV\cite{f:96}.}. 
\item With these $R^0$ and photino masses, the photino relic density
is naturally of the correct order of magnitude to account for the dark
matter of the universe\cite{f:100}.
\item The $R^0$ lifetime $\tau(R^0) {\,\raisebox{-0.13cm}{$\stackrel{\textstyle>}
{\textstyle\sim}$}\,} (10^{-7} - 10^{-10})
(\frac{M_{sq}}{100 {\rm GeV}})^4$ sec\cite{f:99} is long enough that
it is unlikely to have been detected in existing searches\cite{f:95}. 
\item  An ``extra'' pseudoscalar is predicted in the flavor singlet
meson spectrum  at $\sim 1 \frac{1}{2}$ GeV; such a state has been
observed\cite{f:95,f:103}.
\item The ground state $R$-baryon ($uds \tilde{g}$) may be stable and
provide an explanation for cosmic ray events with energies in excess
of the GZK bound\cite{f:104} and anomalous Cygnus X-3 events\cite{f:99}.
\end{itemize}
In earlier papers I have discussed the phenomenology of light
$R$-hadrons and strategies for detecting them.  Here I focus on the
modifications of the squark signature which this scenario implies, as
well as the consequent modifications to chargino signatures if these
can decay to a squark. 

Aside from having spin-0 and larger mass, squarks are produced much
like quarks.  Once produced, squarks decay dominantly via $S_q
\rightarrow q + \tilde{g}$.  The gluino hadronizes forming a jet, due
to the long lifetime of the $R^0$.  This is to be contrasted with the
conventional case of a very heavy and thus short lived gluino, for
which the photino production is prompt and missing energy is a good
signature\cite{f:23}.  When the $R^0$ in the gluino jet finally
decays, the energy carried by the photino is so small that the
conventional missing energy signature is not useful\cite{f:51}.
Existing collider limits do not apply\footnote{Squarks decay directly
to a photino and quark with a branching fraction $Q_{sq}^2
\alpha_{em}/(\frac{4}{3} \alpha_s)$.  Rescaling the UA(1) and Tevatron
collider limits to account for this factor 100 loss in sensitivity,
produces limits inferior to those discussed below from the $Z^0$
hadronic width.}.   

In prinicple, squarks can be reconstructed by pairing jets.  However
experience with $W\rightarrow q \bar{q}$ and $t \rightarrow b q
\bar{q}$ shows that at a hadron collider further constraints will be
necessary to reduce QCD background.  The remainder of this paper is
devoted to establishing a search procedure for squarks when their
predominant decay is to two or more jets.  

Squarks with masses almost $\frac{1}{2} E_{cm}$ can be pair produced
in $e^+ e^-$ collisions.  In a hadron collider they can be produced 
either in pairs from $q \bar{q}$ annihilation and gluon-gluon fusion,
or singly in association with a gluino or (at the $\sim 1\%$ level) a
photino.  At an $e^- p$ collider they are produced singly in
association with a gluino or photino.  Pair produced squarks generate
events containing four or more jets.    

At an $e^+ e^-$ collider, the search for squarks may be complicated by
the presence of an indirect source of squarks which is potentially
even more important than the direct production, namely cascade
production from pair produced charginos.  When gaugino masses vanish
at tree level, the mass of the lighter chargino is less than $M_W$:
$m(\chi^{\pm}) = (\sqrt{\mu^2 + 4 M_W^2 sin 2 \beta} - |\mu|)/2$.
If a $L$-squark (say $S_{uL}$) or the $R$-stop is lighter
than the chargino, major chargino decay modes will be: $\chi^+
\rightarrow S_{uL} \bar{d_L}$, $\chi^+ \rightarrow S_{tR} \bar{b_L}$,
etc.  Above threshold, chargino pairs are copiously produced in $e^+
e^-$ collisions so the cascade chain $e^+ e^- \rightarrow \chi^+
\chi^-$, followed by $\chi^{\pm} \rightarrow S_q q'$ can be the
dominant source of squarks.  Note that the best chargino mass limits
rely on a missing energy signature.  When the branching fraction to
such states is  reduced by competition from $\chi^{\pm} \rightarrow
S_q q'$ the limits on chargino masses are impaired.

The best limit on squark masses prior to LEP running at 130 GeV, if
missing energy is not useful, comes from the determination of the
hadronic width of the $Z^0$.  Neglecting quark masses, $\sigma(e^+ e^-
\rightarrow S_q S_q^*) = \frac{1}{2} \beta^3 \sigma(e^+ e^-
\rightarrow q \bar{q})$  for any given flavor and
chirality\footnote{Although squarks have spin-0, they can be ascribed
chirality because supersymmetry associates them with a quark of
definite chirality and chirality mixing is small for the superpartners
of light-quarks.}.  Therefore production of a $(u_L,~ u_R,~
d_L,~d_R)$-type pair would increase the total hadronic width of the
$Z^0$ by a fraction (0.06, 0.01, 0.09, 0.003)$\beta^3$.  The limit on
``extra'' hadronic width of the $Z^0$ then limits the mass of squarks.
If there are four or more degenerate ``light'' squarks, their mass 
must be greater than $\sim M_Z/2$.  If only a single flavor of squark
is light, this limit is greatly reduced, to ${\,\raisebox{-0.13cm}{$\stackrel{\textstyle<}
{\textstyle\sim}$}\,} 30$ GeV for a
$L$-chiral squark.  Masses of $R$-chiral squarks are comparatively
unconstrained due to their weak coupling to the $Z^0$.  Considering
$e^+ e^- \rightarrow S_q \bar{q} \tilde{g} + S_q^* q \tilde{g}$ and
virtual corrections to $e^+ e^- \rightarrow q \bar{q}$ allows the
degenerate squark limit to be improved to $50-60$
GeV\cite{clavetalsquarks,bhat_deltaR}.  The analysis should be redone
with new $Z^0$ width values and using $\alpha_s(Q)$ determined
assuming a light gluino; the limit should be given as a function
of the number of light squarks of each flavor and chirality. 

Consistency between the observed top mass and its rate in conventional
signatures limits the stop mass to slightly less than the top mass
since otherwise $t \rightarrow S_t + \tilde{g}$ would be the top's
main decay mode.  Limits on isospin-violating radiative corrections to
precisely measured electroweak parameters can be used to constrain the
sbottom-stop splitting, as was done long ago in \cite{bm:custSU2} in a
particular model. 

In the recent LEP run at $E_{cm} = 130-136$ GeV,
ALEPH\footnote{L. Rolandi, Joint CERN Particle Physics Seminar on
First Results from LEP 1.5, Dec. 11, 1995} found 14 events which meet
their 4-jet criteria, when 7.1 events are expected from standard model
physics and less than one 4-jet event is expected from either $hA$ or 
$H^+H^-$ production.  Furthermore, 8 of these 4-jet events have a
total dijet mass of $\sim109$ GeV.  Approximating the statistical
error associated with 7 events by $\pm \sqrt{7}$, the ALEPH data gives
$R_{\ge 4} = 2.0 \pm 0.4$.  If the 7 event excess is averaged over all
four LEP experiments, assuming equal sensitivity and no excess in
other experiments, $R_{\ge 4} = 1.25 \pm 0.1$.  Let us now see whether
such events have a natural interpretation in terms of squark
production.  Even if the ALEPH excess proves ephemeral, the discussion
below provides a vehicle for indicating the types of constraints which
can be brought to bear in every squark search.  We start with the
possibility of direct $S_q S_q^*$ production and then turn to
production via chargino cascade.   

\begin{figure}
\epsfxsize=\hsize
\epsffile{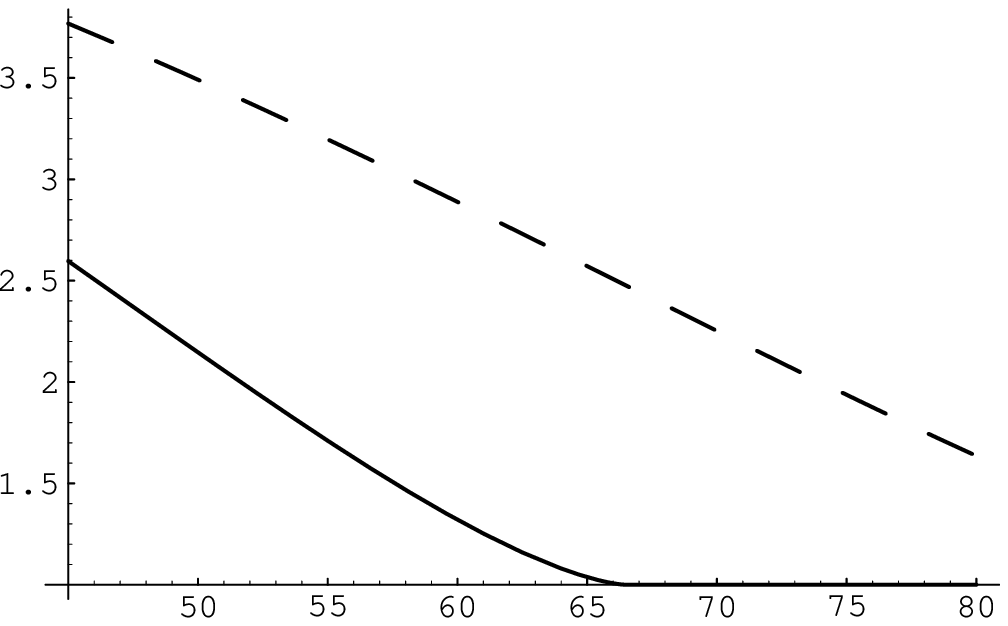}
\caption{$R_{\ge4}$ for degenerate $u,~d,~s,~c$ squarks as a function
of their mass in GeV.  Solid (dashed) curve is for $E_{cm} =
133~(190)$ GeV.}   
\label{sq}
\end{figure}

The most striking feature of squark pair production is the excess
number of events with 4 or more jets.  To be more quantitative, define
$f_{\ge4}$ to be the fraction of ordinary events with four or more
jets, for a given energy and jet-finding algorithm.  In ALEPH's
analysis, $f_{\ge 4} = 0.1$.  For each flavor $i$ define $
r_i(m_i,E) \equiv (\sigma(e^+ e^- \rightarrow S_q^i 
S_q^{i*}))/(\Sigma_i ~\sigma(e^+ e^- \rightarrow q^i \bar{q}^i))$.
Denote the ratio of the actual number of $n_{jet} \ge 4$ events to 
the number expected in the standard model by $R_{\ge4}(E)$. If the
only source of events with $\ge 4$ jets are standard model processes and
$S_q S_q^*$ production, $R_{\ge4}(E) = \frac{\Sigma_i~ r_i(m_i,E) +
f_{\ge4}}{f_{\ge4}}$.  Fig. \ref{sq} shows $R_{\ge 4}(m,E)$ for
$E=133$ and 190 GeV in the illustrative case (called dls) that
$u,~d,~s,~c$ squarks are degenerate.  A LEP measurement of $R_{\ge 4} =
1.25 - 2.4$ at $E_{cm}=133$ GeV implies in the dls case a common
squark mass in the range 47-61 GeV.  This is consistent with the 109
GeV peak in the total dijet mass distribution reported by ALEPH.
Note that $R$-squarks decouple at $E_{cm} \approx 133$ GeV because
their photon and $Z^0$ contributions just cancel.  Thus only
$L$-squarks are probed at this particular energy. Furthermore, at this
energy $U$- and $D$- type squarks are produced equally, if they have
the same $\beta^3$ factor, making it easy to rescale from the dls case
to more complicated squark mass spectra.   

A $\sim 54.5$ GeV chargino and slightly lighter squark can also 
account for the rate of excess dijet events and the peak in total
dijet invariant mass reported by ALEPH.  From a parton point of view,
the chargino cascade mechanism produces 6-jet and not 4-jet events.
However when the chargino-squark mass difference is small, the energy
of each primary quark jet ($q'$ in $\chi \rightarrow S_q q'$) is too
low for it to be distinguished as a separate jet.  Instead, the
particles of the primary jets are associated to the hard jets of the
event.  Thus $\chi^+ \chi^-$ production at this energy would lead to
4-jet events.  In order to account for a peaking of the dijet total
mass at 109 GeV, the chargino mass must be around $54.5$ GeV. 

\begin{figure}
\epsfxsize=\hsize
\epsffile{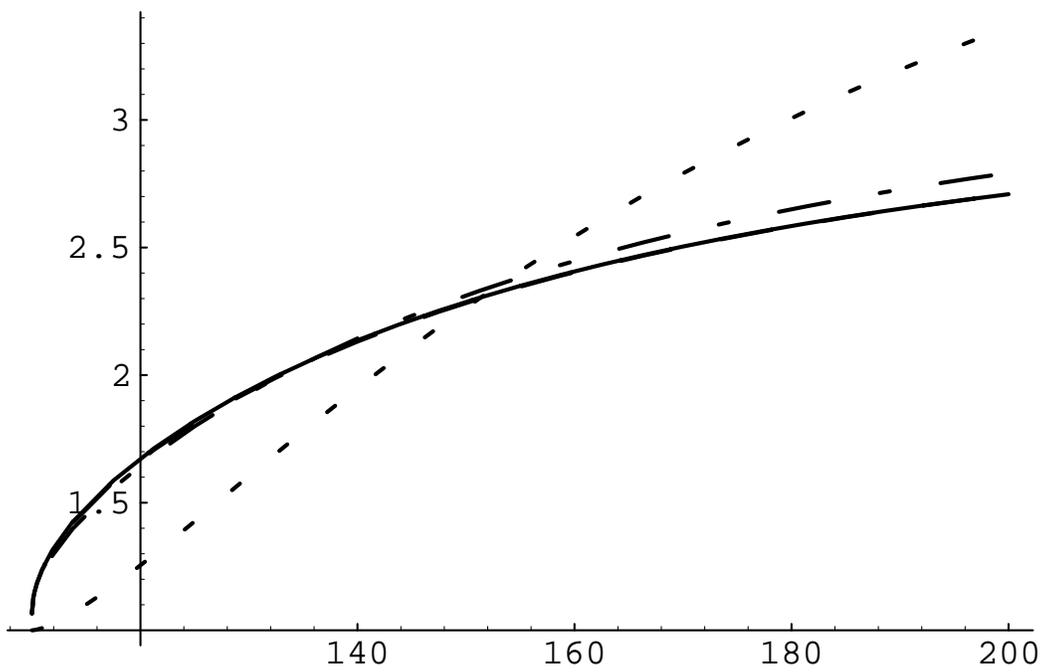}
\caption{$R_{\ge4}$ as a function of $E_{cm}$ for a 55 GeV chargino
with $tan \beta = 1,~ 1.4,$ and 1.8 (solid, dashed and dash-dotted
curves) and for 55 GeV dlsquarks (dotted curve).}  
\label{EdepR4}
\end{figure}

Without tree-level gaugino masses, the masses and mixings of the
charginos, and thus their production cross section, depends only on
$\mu$ and $tan \beta$.  The range of $\mu$ and $tan \beta$
corresponding to a given chargino mass is quite restricted:  e.g., for
$m(\chi^{\pm}) = 55~[65]$ GeV, $tan \beta$ only ranges from 1 at the
maximum allowed value of $\mu$ ( 62 [34] GeV, to 1.8 [1.4] when $\mu =
0$.\footnote{In the MSSM such small $tan \beta$ can conflict with Higgs
mass bounds, however additional scalars are expected in most models and
this is is not a generic problem.}  As for production of any heavy
fermion pair, the threshold dependence of the cross section $\sim
\beta (3 - \beta^2)/2$.  In order to compute the 4-jet event rate in
the chargino cascade mechanism, one needs the branching fraction, $b$,
for $\chi^{\pm} \rightarrow S_q q'$.  This depends on the number and
masses of sneutrinos which are lighter than the chargino.  Fig.
\ref{EdepR4} shows $R_{\ge4}$ for a 55 GeV chargino and one 53 GeV
$L$-squark, as a function of $E_{cm}$ for $tan \beta = 1, ~1.4$, and
1.8 ($\mu$ is fixed by the chargino mass).  $b$ has been fixed to
(0.37,0.38,0.48) for these three values of $tan \beta$, in order that
$R_{\ge 4} = 2$ at 133 GeV.   

It is quite plausible for $b$ to be in this range for a squark mass in
the 50-53 GeV range and sneutrino masses about 5 GeV lighter.  Then
the branching fraction for $\chi^{\pm} \rightarrow S_{\nu} \nu$,
summed over snuetrinos lighter than the chargino, is $(1-b)$.  At
$E_{cm} = 133$ GeV, $b$ values giving 7 excess 4-jet events imply
14-24 events with one chargino decaying to $S_q q$ and the other to
$S_{\nu} \nu$, and 7-14 events with both charginos decaying to
$S_{\nu} \nu$.  The former produces events with 2 jets, a soft charged
lepton ($E \sim m(\chi^{\pm}) - m(S_{\nu})$), and the decay products
of the sneutrino; the latter produces events with two soft leptons and
decay products of two sneutrinos. It should be possible to verify or
exclude the chargino cascade scenario by searching for these alternate
decay modes.

Now let us turn to the issue of deciding when an excess of events
with $n_{jets} \ge 4$ can be attributed to direct production of $S_q
S_q^*$.  Three techniques are particularly useful:  (i) dijets
from $S_q S_q^*$ production should have equal masses, (ii) the angular
distribution of jet clusters should $\sim sin^2 \theta$ and $1 + cos^2
\theta$ for squarks and charginos, and (iii) gluino-jet tagging. 

Gauge interactions (including their SUSY-transforms involving
gauginos) conserve chirality.  Moreover the absence of flavor-changing
neutral currents implies that gauge interactions of squarks are
flavor-diagonal to high accuracy.  Thus when a squark and antisquark
pair is produced in $e^+ e^-$ or hadron colliders, their flavor and
chirality can be taken to be the same until the sample of squarks is
large enough to study rare phenomena.  Furthermore, the {\it mixing}
between eigenstates of chirality for a given flavor squark is small in
this scenario\footnote{Because trilinear squark-Higgs couplings are
absent, the mixing between chirality eigenstates for squark flavor $q$
is $ \frac{\mu m_q}{{M_q}^2}$ times $cot \beta$ ($tan \beta$) for
charge 2/3 (-1/3) squarks respectively.}, except for the stops.  Thus
the {\it dijets from a directly produced squark pair have equal masses
when the jets are correctly paired.}.  This is a crucial point. Since
the various squark flavors need not be degenerate, the dijet invariant
mass spectrum may be messy, with nearby overlapping peaks or
enhancements.  Nonetheless, a clear signal is possible since correct
pairing of jets always leads to a vanishing {\it difference} of dijet
invariant masses.  Henceforth jets are always taken to be paired so
the dijet mass difference is minimized.  When the squarks are decay
products of charginos they need not be identical if more than one
squark is lighter than the chargino.   Nonetheless the mass splitting
may not be large\footnote{Aside from stops, squarks produced from
chargino decay are left-chiral because their gauge coupling to the
wino component of the chargino is much larger than their super-Yukawa
coupling to the higgsino component.  The splitting between $S_{UL}$
and $S_{DL}$ is determined as the soft SUSY-breaking scalar mass terms
respect the electroweak gauge symmetry.  In obvious notation, $ M_U -
M_D =  (cos 2 \beta (1 - sin^2 \theta_W)M_{Z^0}^2 + m_U^2 - m_D^2)/
2M_Q$, where $M_Q$ is their average mass.}.   

Unfortunately, the prediction that directly produced squarks and
antisquarks are mass degenerate on an event by event basis is not a
very useful tool near threshold.  ALEPH modeled the distribution of
dijet mass difference at $E_{cm} = 135$ GeV resulting from two 55 GeV
particles each decaying to $q \bar{q}$ and found the peak in
reconstructed dijet mass difference to be $\sim 15$ GeV fwhm.
Furthermore, requiring the minimum dijet mass difference to be less
than 20 GeV only reduces the number of events expected in the standard
model from 8.6 to 7.1$.^5$  This implies that at $E_{cm} = 135$ GeV,
80\% of standard model events have a dijet mass difference less than
20 GeV, when jets are paired so as to minimize the dijet mass
difference and the other ALEPH cuts are satisfied. Thus while the data
is consistent with equal mass squarks, this is not a very stringent test
of the hypothesis.      

If both direct and cascade production of squarks are important and the
squark and charginos are close in mass, the event properties near
chargino threshold are complicated.  The particles of the soft primary
quark jets in chargino decays would be associated in a more-or-less
random way with the four hard jets.  This would tend to broaden the
dijet invariant mass distribution compared to direct squark pair
production, although the average invariant mass of the dijet remains centered
on the chargino mass to leading approximation.  In the ALEPH anaylsis,
the cut on min($m_i + m_j)>10$ GeV removed only 2 events from the data
but 5 from the monte carlo of the SM prediction.  This may be a hint
of chargino cascade, because when the particles of the very soft
primary quark jets are associated with the 4 hard jets, the invariant
mass of the resultant jets increases.  

Spin-0 particles produced in $e^+ e^-$ scattering through the spin one
photon or $Z^0$ have a $sin^2 \theta$ angular distribution.  If the
events with total dijet mass $\sim 109$ GeV were due to the decay of
directly produced squark pairs, taking these events alone should
produce an angular distribution $\sim sin^2 \theta$.  The remaining
events (comprised of $q \bar{q} g g$ in the absence of chargino
cascade) should be produced according to the standard model and thus
have a different characteristic angular dependence. 

Since the total momentum of the chargino is the vector sum of the
three-momentum of the squark jet and the unidentified soft quark jet,
there is no direct relation between the 3-momentum attributed to the
dijet and the actual chargino 3-momentum.  Thus the chargino angular
distribution cannot be determined near threshold.  Fortunately, this
situation improves as $E_{cm}$ is increased.  When the jets from
chargino decay are collimated by a Lorentz boost, the ambiguity of associating
particles with the correct chargino is reduced and the invariant mass
and three-moemntum of the tri-jet systems should reconstruct to the
chargino mass and 3-momentum, even when the three jets are not
independently resolved.  The angular distribution of the events in the
mass peak should in this case $\sim (1 + cos^2 \theta) $. 

Squarks are pair produced in flavor eigenstates, to a good
approximation, so that $S_q S_q^*$ events should contain two gluino
jets and two jets of the same flavor, e.g., $b$ and $\bar{b}$ or $c$ and
$\bar{c}$.  There will often be additional gluon jets since with
typical jet definitions (e.g., $y_{cut} = 0.01$), 40\% of the hadronic
$Z^0$ decays have $\ge 3$ jet final states.  The hadronization of
gluino jets will nearly always produce an $R^0$\cite{f:95} which
ultimately decays to a photino which escapes. $R^0$-tagging could
provide confirmation of the $S_q S_q^*$ origin of the excess $\ge
4$-jet events.  The $R^0$'s decay to a photino and a small number of
pions\cite{f:102}.  The photino typically has a momentum transverse to
the $R^0$ direction of $\sim 0.4 - 0.8$ GeV\cite{f:102}, depending on
the relative mass of $R^0$ and $\tilde{\gamma}$.  The average momentum
fraction of an $R^0$ with respect to its jet, $x_R$, can be determined
in a Monte Carlo or other model of jet fragmentation\footnote{The
$x_{\tilde{g}}$ distribution and invariant mass of gluino jets was
estimated in \cite{deR_P}.}, or taken by analogy from, say, charm
fragmentation. If the lifetime of the $R^0$ is short compared to the
transit time of the calorimeter and its decay is two-body, the photino
will have a momentum along the jet ranging up to $ x_R \frac{
M_{sq}}{2}$.  On the other hand, if the $R^0$ lifetime is long 
enough that it loses its kinetic energy in the calorimeter before
decaying, the momentum along the jet axis carried away by the photino
will be nearly imperceptible.   

As can be seen from Fig. \ref{EdepR4}, with improved statistics and higher
energy LEP measurement of $R_{\ge 4}$ will provide a powerful tool to
support or exclude the  hypothesis that squarks are being produced.
Mapping the $\ge4$-jet energy dependence should allow cascade and
direct production to be distinguished.  If charginos are being
produced, the threshold dependence of $R_{\ge4}$ allows their mass to
be found.  If a chargino with mass less than $m_W$ is eventually
excluded, SUSY-breaking which does not produce tree level gaugino
masses will be ruled out unless $\mu$ is much larger than has been
considered plausible up to now\cite{f:96}.  If on the other hand a
chargino but not squarks are found below the $W$, the search for
squarks in $\ge 4$-jet events should be pursued at higher energy.  The
range of LEP can be extended somewhat by allowing one member of the
squark pair to be off-shell. Then only one pair of jets will
reconstruct to a definite invariant mass and the signal will be less
clear.  

At a hadron collider the background is more severe, but the
signal-to-noise improves with increasing squark mass.  Chargino
cascade is not a significant source of squarks in a hadron collision,
so with good resolution a peak should be found in the mass difference
of dijets, with the flavor of two of the jets being the same.  At a
cost of a factor of 100 in rate, one could trigger on events in which  
a squark decays to quark and photino.  This produces events with
missing energy, one dijet and one or more additional jets.  These
events would contain, ignoring mass differences, equal numbers of $L$-
and $R$-squarks with a 4-1 enhancement of events with $U$-squarks.  The
mass-splitting between $L$- and $R$-squarks of the same flavor is not
determined by $tan \beta$ as it is within an SU(2) doublet, since
squarks in different SU(2)xU(1) representations can have different
soft SUSY-breaking masses.  If at low energy the SUSY-breaking
contribution to scalar masses is universal, the $S_{uL} - S_{uR}$ mass
splitting is $cos 2 \beta  (\frac{1}{2} - \frac{4}{3} sin^2 \theta_W) 
\frac{m_Z^2}{2 \bar{M}}$. This is less than 2 GeV, for $m(\chi^{\pm}
\ge 70$ GeV and $ M_sq \ge 100$ GeV.  In this case events with missing
energy and $\ge 3$ jets would show a peaking in dijet mass, even
though the squarks are not precisely degenerate. 

In $e^- p$ collisions a produced squark is 4-8 times as likely to
be a $u$-squark as a $d$-squark depending on the $x$ regime of the
collision: a factor of 4 from the quark charge-squared and a factor
of $1-2$ from the relative probability that the initial parton
is a $u$ versus $d$ quark.  Ignoring therefore the production of
$d$-squarks, the probabilitiy that squark production leads to a prompt
photino is 4\%.  Half the time the photino is associated with a dijet
which reconstructs to the $u$-squark mass; the other half of the time the
two jets accompanying it have no particular relation as one came from
the decaying squark and the other was the primary gluino associated
with the squark production.  

To recapitulate, the signatures of squarks and charginos have been
discussed when SUSY breaking does not produce tree-level gaugino
masses or scalar trilinear couplings.  In this case a chargino must be
lighter than the $W$ and the gluino is light and hadronizes.  If a
$L$-squark is lighter than the chargino, the primary decay of the
chargino is to squark + quark and the usual signatures relying on the
decay $\chi^{\pm} \rightarrow \chi^0 f \bar{f'}$ are diminished in
utility.  99\% of pair-produced squarks produce events with four or
more jets and little missing energy.  Emphasis was placed on features
of events with four or more jets which are characteristic of 
$S_q S_q^*$ production or (possibly relevant at LEP) of chargino
production and decay to squarks.  The energy dependence of events with
4 or more jets is a powerful tool to establish the existance of a
signal and to discriminate between cascade and direct squark
production.  Properties of gluino-containing jets which could be
helpful in discriminating them from quark or gluon jets are discussed. 

The excess of 4-jet events reported recently by ALEPH$^5$, if
confirmed by other experiments and higher statistics, could be 
circumstantial evidence that at least one $L$-squark has a mass ${\,\raisebox{-0.13cm}{$\stackrel{\textstyle<}
{\textstyle\sim}$}\,}
55$ GeV and decays to quark and hadronizing gluino.  The number of 4-jet
events in their total-dijet-mass peak at 109 GeV is consistent with
direct production of two generations of $L$-squarks with mass 55 GeV,
or production of a 55 GeV chargino which decays to a squark and quark.
In the latter case, events with a pair of jets and decay products of
sneutrinos are expected.  Higher energy running can easily exclude or
confirm these possibilities.  Careful study of 4-jet events should be
a standard part of squark search techniques until a light, hadronizing
gluino has been excluded.     

{\bf Acknowledgements:}  I have benefited from discussing these matters
with L. Clavelli, J. Conway, P. Janot, S. Lammel and R. Rattazzi.




\end{document}